\documentclass[aps,prl,reprint,noeprint,superscriptaddress]{revtex4-2}

\usepackage{graphicx}
\usepackage{url}
\usepackage{amsmath}
\usepackage{amssymb}
\usepackage{xfrac}
\usepackage{dsfont}
\usepackage{braket}
\usepackage{bm}
\usepackage{bbm}
\usepackage{color}

\newcommand{\afflkb}{Laboratoire Kastler Brossel, ENS-Universit\'{e} PSL, CNRS, Sorbonne Universit\'{e}, Coll\`{e}ge de France, 24 rue Lhomond, 75005 Paris, France}

\begin{document}

\title{\emph{In Situ} Thermometry of Fermionic Cold-Atom Quantum Wires}

\author{Cl\'{e}ment~De~Daniloff}
\affiliation{\afflkb}
\author{Marin~Tharrault}
\affiliation{\afflkb}
\author{C\'{e}dric~Enesa}
\affiliation{\afflkb}
\author{Christophe~Salomon}
\affiliation{\afflkb}
\author{Fr\'{e}d\'{e}ric~Chevy}
\affiliation{\afflkb}
\author{Thomas~Reimann}
\affiliation{\afflkb}
\author{Julian~Struck}
\email[Corresponding author:~]{julian.struck@lkb.ens.fr}
\affiliation{\afflkb}

\begin{abstract}
We study ensembles of fermionic cold-atom quantum wires with tunable transverse mode population and single-wire resolution. From \emph{in situ} density profiles, we determine the temperature of the atomic wires in the weakly interacting limit and reconstruct the underlying potential landscape. By varying atom number and temperature, we control the occupation of the transverse modes and study the 1D-3D crossover. In the 1D limit, we observe an increase of the reduced temperature $T/T_{F}$ at nearly constant entropy per particle $S/N k_{B}$. The ability to probe individual atomic wires \emph{in situ} paves the way to quantitatively study equilibrium and transport properties of strongly interacting 1D Fermi gases.
\end{abstract}

\maketitle

The 1D world represents an exotic realm of many-body physics. Quantum and thermal fluctuations are enhanced and the dimensional constraint on the motion of particles strongly increases the impact of interactions \cite{Giamarchi2004}. A paradigm for the resulting unconventional behavior is the complete collectivization of elementary excitations in gapless 1D systems, known as Tomonaga-Luttinger liquids (TLLs) \cite{Tomonaga1950,Luttinger1963,Haldane1981a}. Signatures for TLLs and other characteristic 1D states have been observed in a variety of solid-state systems, including organic conductors \cite{Schwartz1998,Giamarchi2008}, carbon nanotubes \cite{Bockrath1999,Ishii2003}, semiconductor wires \cite{Yacoby1996,Auslaender2005,Jompol2009}, antiferromagnetic spin chains \cite{Lake2005}, metallic chains \cite{Segovia1999} and edge modes of integer and fractional quantum hall states \cite{Chang2003}. However, these materials are complex and typically feature uncontrolled interdimensional couplings, rendering quantitative studies difficult.

Ultracold atomic gases provide a complementary approach to low-dimensional many-body systems \cite{Bloch2008b,Pitaevskii2016,Krinner2017}. Their motional degrees of freedom can be tailored precisely via optical or magnetic potentials, and confinement-induced resonances provide a means to tune the sign and strength of interactions \cite{Olshanii1998,Bergeman2003,Moritz2005,Haller2010}. This high degree of controllability makes 1D Fermi gases promising candidates for the observation of elusive phenomena, such as itinerant ferromagnetism \cite{Gharashi2013,Cui2014,Jiang2016}, Fulde-Ferrell-Larkin-Ovchinnikov (FFLO) pairing \cite{Orso2007,Hu2007} or Majorana edge states \cite{Jiang2011,Liu2012}. So far, only a handful of experiments investigated the properties of 1D bulk Fermi gases \cite{Moritz2005,Liao2010,Pagano2014,Revelle2016,Yang2018,Chang2020}, including pioneering works on the control of interactions \cite{Moritz2005,Chang2020} and the effect of spin imbalance in such systems \cite{Liao2010,Revelle2016}. These experiments, however, suffered from addressing arrays of wires stacked along two spatial directions with varying atom number, yielding ensemble-averaged measurements. While this drawback did not necessarily bias all previous studies, it certainly prevented progress towards the quantitative understanding of many-body problems in 1D. Indeed, as the thermodynamic state directly depends on the density, these ensemble-averages cover extended regions of the phase diagram, which severely complicates their interpretation and potentially obscures signatures of elusive states. On an even more fundamental level, ensemble averages pose a problem for the observation of critical behavior or states that are characterized by spontaneous pattern formation, e.g., magnetic domains.

In this Letter, we report on the preparation, detection and thermodynamic characterization of individual fermionic cold-atom quantum wires in the 1D regime and 1D-3D crossover. Our approach relies on the selective loading of a single plane of a 2D optical lattice and high-resolution imaging of the resulting single row of atomic wires. This strategy allows us to circumvent line-of-sight averaging in the absorption images and to directly access the density distribution of each wire. The ability to resolve 1D density profiles \emph{in situ} and perform local thermometry represents the main result of this work. In addition, we precisely characterize the trapping potential, which is a crucial  prerequisite for further thermodynamic studies of strongly interacting 1D Fermi gases \cite{Ho2009,Nascimbene2010,Navon2010,Yefsah2011,Ku2012,Fenech2016,Boettcher2016}.


\begin{figure*}
\centering
\includegraphics[width=\textwidth]{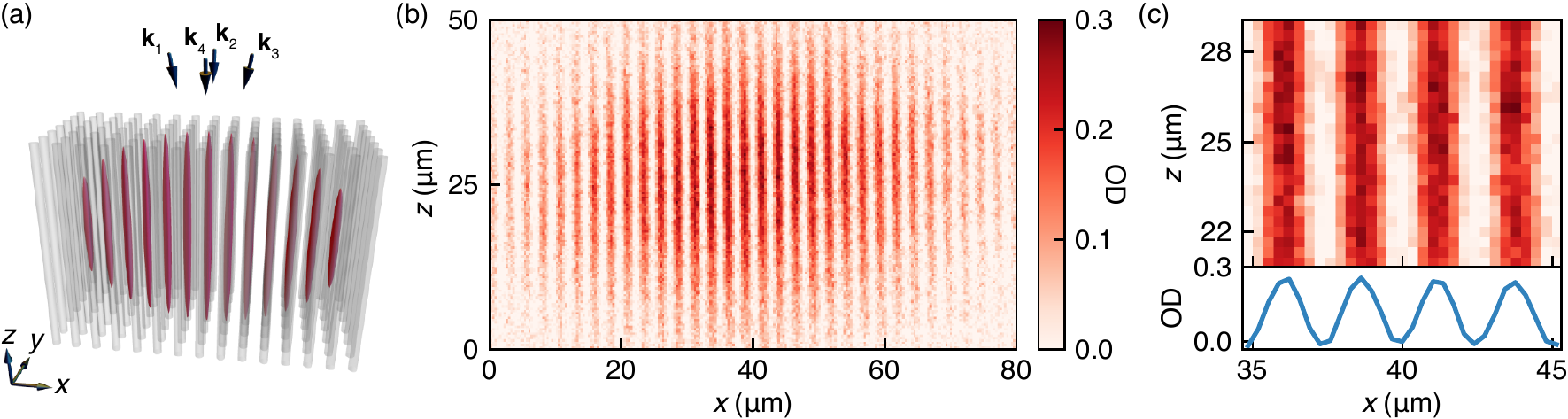}
\caption{Individually resolved fermionic quantum wires. (a) Array of tube traps created by superimposing two optical lattices along the $x$- and $y$-direction. The optical lattices are formed by the pairwise interference of the beams under small angles ($\measuredangle (\mathbf{k}_{1},\mathbf{k}_{3}) \approx \measuredangle (\mathbf{k}_{2},\mathbf{k}_{4}) \approx 24^{\circ}$), resulting in a lattice spacing of $d \approx 2.6 \, \mathrm{\mu m}$. Here, $\mathbf{k}_{i}$ denotes the wave vector of the $i$th beam. Only a single row of tube traps is populated with atoms. (b) 32 averaged \emph{in situ} absorption images of Fermi gases in the 1D regime. Shown is the optical density (OD). (c) Central region of interest of (b) and the corresponding integrated line profile. For a single absorption image see \cite{Note1}.}
\label{fig:M1}
\end{figure*}

Experimentally, reaching the 1D regime with an atomic gas requires a tight transverse confinement. The occupation of the energetically lowest transverse mode must be predominant and excitations have to be strongly suppressed. This implies that the transverse quantum of energy has to be large compared to the energy scales of the gas, i.e., the Fermi energy $E_{F}$ and thermal energy $k_{B}T$, where $k_{B}$ is the Boltzmann constant and $T$ the temperature. We employ a large-spacing 2D optical lattice to create an array of independent tube-shaped traps with the potential \mbox{$V(\rho,z) = m{\omega_{\perp}}^{2}{\rho}^{2}/2 + V_{\parallel}(z)$}, where $\rho^2 = x^2+y^2$, $m$ is the mass of the atoms and $V_{\parallel}(z) = m{\omega_{\parallel}}^{2}{z}^{2}/2 + \mathcal{O}(z^3)$ the axial potential. Given this potential, the 1D limit is expressed as $k_{B}T\ll \hbar \omega_{\perp}$ and $E_{F} \ll \hbar \omega_{\perp}$. The transverse and axial trap frequencies are $\omega_{\perp}/2\pi \approx 17\,\mathrm{kHz}$ and $\omega_{\parallel}/2\pi \approx 96\,\mathrm{Hz}$, which corresponds to a ratio of $\omega_{\perp}/\omega_{\parallel}=177$ \footnote{See attached supplemental material for more information regarding the large-spacing optical lattice, absorption imaging, optical pumping tomography, measurement of the axial frequency, the equation of state fitting procedure, transverse mode occupation and the isentropic 1D-3D relation of the reduced temperature for the noninteracting Fermi gas.}.
The 2D optical lattice is composed of two orthogonally intersecting standing waves that are each created by interfering a pair of laser beams under a small angle [see Fig. \ref{fig:M1}(a)]. This results in a lattice constant of $d = 2.6\,\mathrm{\mu m}= 2.4\lambda$, where $\lambda=1064\,\mathrm{nm}$ is the wavelength of the laser beams. The large lattice spacing renders tunneling between the tube traps negligible. We measure the \emph{in situ} density distribution of the atomic wires in the tubes through high-resolution absorption imaging [see Fig. \ref{fig:M1}(b)] \cite{Note1}. Our imaging resolution, as defined by the Rayleigh criterion, is $1.3\,\mathrm{\mu m}$. This is twice lower than the lattice spacing. Therefore, individual atomic wires are fully resolved [see Figs. \ref{fig:M1}(c) and (d)]. Crucially, to avoid line-of-sight integration along the imaging axis, only a single row of tube traps is populated with atoms [see Fig. \ref{fig:M1}(a)].

\begin{figure}
\centering
\includegraphics[width=8.6cm]{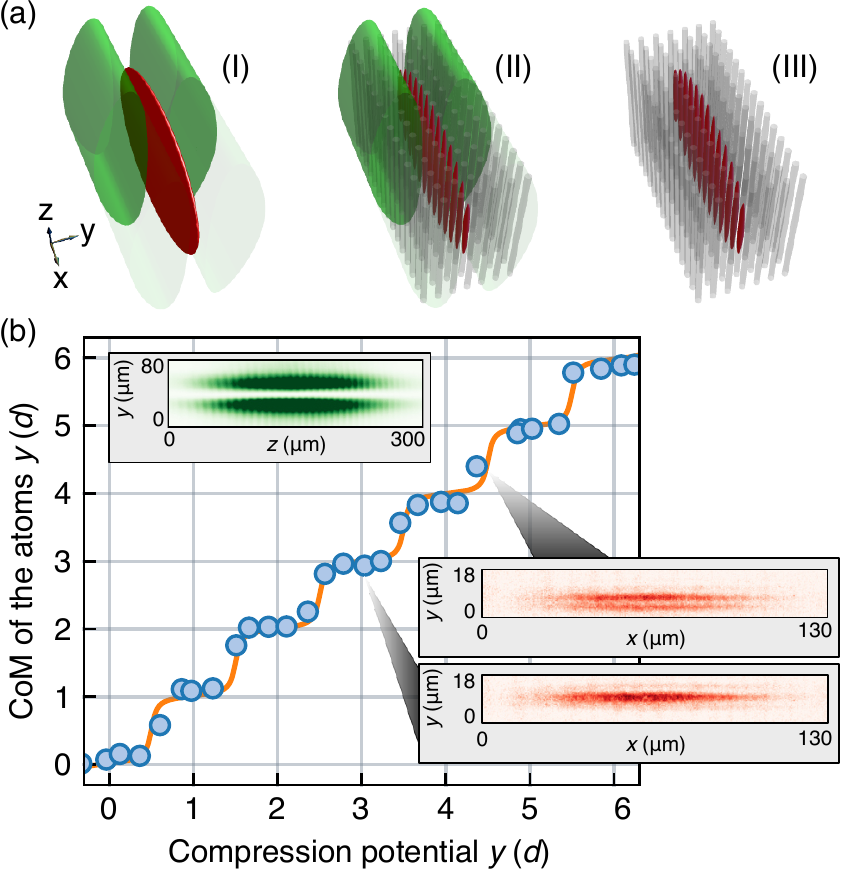}
\caption{Lattice loading procedure. (a) (I) First, the gas (red) is strongly compressed along the $y$-direction by a repulsive $\mathrm{TEM}_{01}$-like optical potential (green). (II) Second, the optical lattice is ramped up, populating only a single row of the array of tube traps. (III) Finally, the compression potential is removed. (b) Center of mass (CoM) of the atoms in the lattice as a function of the compression beam center along the $y$-axis in units of the lattice constant $d$. The solid line is a guide to the eye. The top inset shows the intensity profile of the compression beam. The insets on the right represent characteristic images for single- and double-row loading.}  
\label{fig:M2}
\end{figure}

Our experiments are conducted with a balanced mixture of $^{40}\mathrm{K}$ atoms in the two energetically lowest hyperfine states, representing a pseudospin-$1/2$ system. The interstate contact interactions are controlled via an $s$-wave Feshbach resonance at $202.10(7)\,\mathrm{G}$ \cite{Regal2004}. Initially, the quantum degenerate gas is strongly compressed in a crossed optical dipole trap by superimposing a repulsive $\mathrm{TEM}_{10}$-like optical potential [see Fig. \ref{fig:M2}(a)] \cite{Smith2005,Rath2010,Note1}. At full compression beam intensity, the pancake-shaped potential is characterized by trap frequency ratios of $\omega_{y}/\omega_{x}=18$ and $\omega_{z}/\omega_{x}=2.6$. The position of the pancake potential determines which tube traps are selectively loaded when ramping up the optical lattice. Once the lattice depth is sufficient to inhibit tunneling, the compression potential is removed [see Fig. \ref{fig:M2}(a)].

To verify the loading of a single row of tubes, we image the atomic cloud along the vertical direction ($z$-axis). The large axial spread of the atoms in each tube trap poses a problem for high-resolution imaging given its shallow depth of field. We mitigate this issue by using a tomographic optical pumping scheme that transfers atoms outside the central region into undetected hyperfine states \cite{Andrews1997a,Ku2014,Note1}. While it is impossible to resolve individual atomic wires along this imaging direction, we can clearly distinguish between the loading of a single and double row of the lattice [see insets of Fig. \ref{fig:M2}(b)]. Translating the compression beam orthogonally with respect to the tube traps, we observe a sharp steplike structure when tracking the center of mass of the cloud [see Fig. \ref{fig:M2}(b)]. The plateaus correspond to the population of different rows. This indicates the robustness of the loading procedure against misalignment of the compression beam and small phase drifts of the optical lattice. In addition, long-term drifts of the lattice are actively compensated by readjusting the phase at the beginning of each sequence \cite{Note1}.

\begin{figure}
\centering
\includegraphics[width=8.6cm]{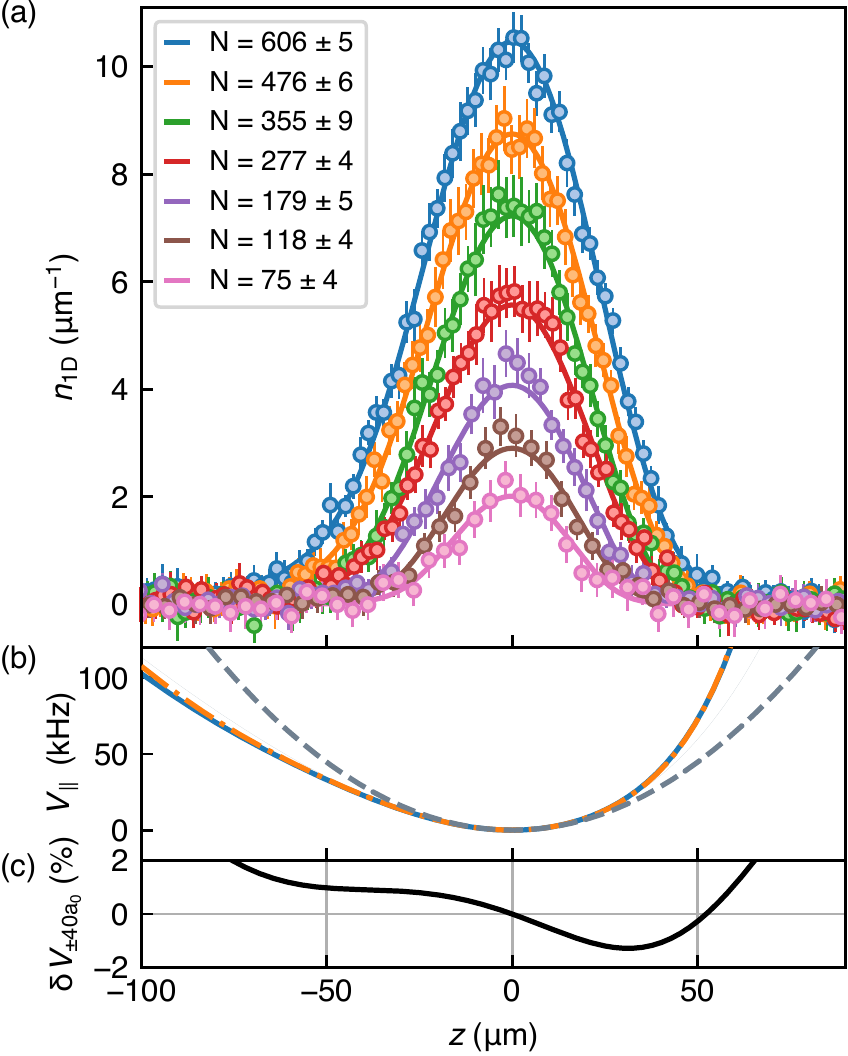}
\caption{(a) Averaged 1D density profiles for individual tubes with a total atom number $N$ per spin state. The solid lines represent fits of the noninteracting equation of state, with $T$ (from top to bottom): $0.51(1)\,\mathrm{\mu K}$, $0.44(2)\,\mathrm{\mu K}$, $0.31(2)\,\mathrm{\mu K}$, $0.41(2)\,\mathrm{\mu K}$, $0.29(2)\,\mathrm{\mu K}$, $0.27(2)\,\mathrm{\mu K}$, and $0.25(3)\,\mathrm{\mu K}$. (b) Reconstructed axial potential for $a=-40\,a_{0}$ (orange dashed-dotted line) and $a=+40\,a_{0}$ (blue solid line). The harmonic part (gregrayy dashed line) of the potential has been independently measured \cite{Note1}. (c) Relative difference $\delta V_{\pm 40a_{0}}(z)= (V_{\parallel,-40a_{0}}(z) - V_{\parallel,+40a_{0}}(z))/V_{\parallel,-40a_{0}}(z)$ between the reconstructed potentials for attractive and repulsive interactions.}  
\label{fig:M3}
\end{figure}

Now, we turn to the thermodynamic analysis of \emph{in situ} density profiles. By splitting the absorption images [see Fig. \ref{fig:M1}(b)] into sections containing only a single atomic wire and integrating over the $x$-axis, we obtain the individual 1D density distributions $n_{\mathrm{1D}}(z)$ per spin state. To improve the signal to noise ratio, we average profiles with comparable atom numbers prepared under the same experimental conditions [see Fig. \ref{fig:M3}(a)]. For the thermometry, the 3D scattering length is reduced to $|a|=40\,a_0$, which is still sufficient to establish thermal equilibrium. Here, $\,a_0$ is the Bohr radius. To determine the temperature $T$ and chemical potential $\mu$, we use the equation of state of the noninteracting Fermi gas
\begin{equation}
	n_{\mathrm{1D}}(\mu,T) = -\frac{1}{\lambda_{T}} \sum_{s=0}^{\infty} (s+1)\, \mathrm{Li}_{\frac{1}{2}}(-f_{s}(\mu,T)),
	\label{eqn:DensityEOS}
\end{equation}
where the sum accounts for the transverse modes with the energy $E_{s}=\hbar \omega_{\perp}(s+1)$ and degeneracy $s+1$.
Here, $\mathrm{Li}_{n}(x)$ is the $n$th order polylogarithm, $\lambda_{\mathrm{T}}=h/\sqrt{2 \pi m k_{B} T}$ the thermal de Broglie wavelength and $f_{s}(\mu,T)=\exp(\mu/k_{B} T)\exp(-s \hbar \omega_{\perp}/k_{B} T)$ the fugacity of the $s$th transverse mode. Our assumption of negligible interaction effects in the thermodynamic analysis is not justified \emph{a priori}. In fact, in the quantum degenerate 1D regime, the ratio of interaction and kinetic energy ($E_{\mathrm{int}}/E_{\mathrm{kin}} \sim 1/n_{\mathrm{1D}}$) diverges in the low-density limit. This leads to a nontrivial competition of kinetic energy and interaction effects in the wings of the atomic wires, where the density and degeneracy drop simultaneously. To estimate the influence of interactions on the density profiles, we perform an \emph{a posteriori} consistency check of our analysis for repulsive and attractive interactions ($a = \pm 40\,a_{0}$).

Applying the local density approximation $\mu(z)=\mu_{0} - V_{\parallel}(z)$ to the weakly confined axial direction allows us to fit the equation of state [Eq. \eqref{eqn:DensityEOS}] to the measured 1D density distributions. By fitting a set of profiles with varying temperatures and atom numbers [see Fig. \ref{fig:M3}(a)], we extract $\mu$ and $T$ as independent parameters for each profile and the reconstructed axial potential as a shared parameter for the entire set \cite{Salces-Carcoba2018a}. More precisely, the anharmonic part of the potential is modeled by a higher-order polynomial while the harmonic part is fixed through an independent measurement \cite{Note1}. The asymmetry and anharmonicity of the potential [see Fig. \ref{fig:M3}(b)] stem from the gravitational force along the $z$-direction as well as the Gaussian profiles of the laser beams forming the optical lattice and crossed optical dipole trap. We observe no significant variation of the axial potential across the 18 central tube traps, which are selected for the data analysis. Comparing reconstructed potentials from sets with attractive and repulsive interactions reveals only minor differences [see Fig. \ref{fig:M3}(c)] and thus validates the use of the noninteracting equation of state.

\begin{figure}
\centering
\includegraphics[width=8.6cm]{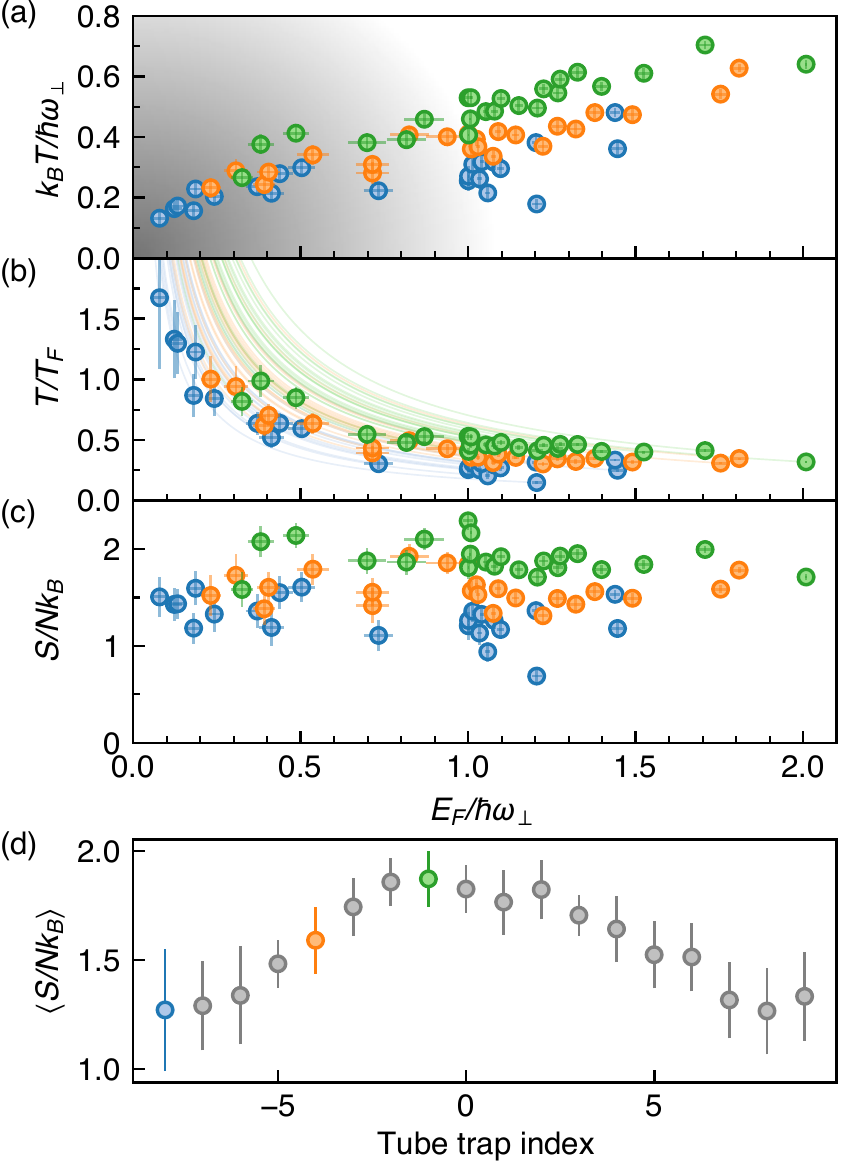}
\caption{Thermodynamics of the 1D-3D crossover. The data points in (b), (c) and (d) correspond to the center of each tube trap ($z=0$), where the Fermi energy is the highest. Green, orange and blue colors represent a selection of three characteristic tubes whose indices are indicated in (d). Tube trap indices label the 18 selected central tubes from left to right along the x-axis. (a) The temperature of individual atomic wires normalized by the transverse frequency $k_{B}T / \hbar \omega_{\perp}$ and (b) the reduced temperature $T/T_{F}$. The gray gradient in (a) depicts the 1D limit. The solid lines in (b) that emanate from the data points represent the continuous change in $T/T_{F}$ and $E_{F}/(\hbar \omega_{\perp})$ within each atomic wire according to the local density approximation. (c) The entropy per particle $S/Nk_B$. (d) Mean value of the entropy per particle obtained by averaging the data of (c) for the green, orange and blue points. The gray points represent the other 15 tubes traps, which were omitted in (a), (b) and (c).}  
\label{fig:M4}
\end{figure}

With the thermometry at hand, the 1D-3D crossover, driven by the gradual occupation of transverse modes, can be precisely characterized. We obtain local dimensionless quantities by normalizing the relevant energy scales with the Fermi energy, which is determined by the implicit equation
\begin{equation}
	n_{\mathrm{1D}} = \sqrt{\frac{8 m E_{F}}{h^2}} \sum_{s=0}^{\left \lfloor \frac{E_{F}}{\hbar \omega_{\perp}}\right \rfloor} (s+1)\,\sqrt{1-s\frac{\hbar \omega_{\perp}}{E_{F}}},
	\label{eqn:FermiEnergy}
\end{equation}
where $\lfloor x \rfloor$ denotes the floor function. This equation simplifies to $E_{F}= h^2 n_{\mathrm{1D}}^2 / 8m$ in the 1D regime. The population of transverse modes can either be caused by thermal excitations or arise as a consequence of Pauli blocking. By changing the initial evaporation parameters in the crossed optical dipole trap $(T/T_{F} \approx 0.15 - 0.25)$ prior to loading the atoms into the lattice, we vary the final atom number and temperature in the tubes. This way, we control the transverse mode population and can reach the deep 1D regime with $k_{B} T \lesssim 0.2\, \hbar \omega_{\perp}$ and $E_{F} \lesssim 0.2 \, \hbar \omega_{\perp}$ [see Fig. \ref{fig:M4}(a)] \cite{Note1}. The temperature spread across different atomic wires is yet another sign that tunneling between the tube traps is strongly suppressed, which leads to an early thermal decoupling during the lattice loading procedure.

Previous 1D Fermi gas experiments inferred estimates of the degeneracy based on thermometry of the initial 3D cloud before loading the optical lattice. These estimates rely on the assumption of an isentropic loading procedure, which is questionable due to various technical heating processes, e.g, laser intensity noise, and the suppression of thermalizing two-body collisions in the 1D regime \cite{Kinoshita2006,Mazets2008,Gring2012}. More fundamentally, the temperature and reduced temperature $T/T_{F}$ are not conserved in the isentropic 1D-3D crossover \cite{Note1}. The corresponding isentropic 1D-3D relations depend on interactions and are, in general, not known. With our approach, we do not require any knowledge about the loading process of the tube traps and instead determine \emph{in situ} the temperature, local $T/T_{F}$ and local entropy per particle
\begin{equation}
	\frac{S}{N k_{B}} = \frac{\sum_{s=0}^{\infty} (s+1)\left(\frac{3}{2} \mathrm{Li}_{\frac{3}{2}}(-f_{s}) - \mathrm{ln}(f_{s}) \mathrm{Li}_{\frac{1}{2}}(-f_{s}) \right)}{\sum_{s=0}^{\infty} (s+1)\mathrm{Li}_{\frac{1}{2}}(-f_{s})}.
	\label{eqn:Entropy}
\end{equation}
We observe that $S/N k_B$ stays nearly constant in the entire crossover region, whereas $T/T_{F}$ displays a sudden increase in the low atom number and temperature limit [see Fig. \ref{fig:M4}(b),(c)]. This is a clear signature of the 1D regime, where the equation of state is strongly altered with respect to the 3D case. Note that, compared to the 3D ideal Fermi gas, the role of quantum statistics in 1D remains important at higher values of $T/T_F$. For instance, at $T/T_F=1.6$ (the highest $T/T_F$ in Fig. \ref{fig:M4}(b)), Maxwell-Boltzmann statistics overestimates the density of the noninteracting gas by $55\%$ in 1D, versus $12\%$ in 3D.

Interestingly, we observe that the entropy per particle varies between the tubes, with an increase towards the center where the density is highest [see Fig. \ref{fig:M4}(d)]. This may be evidence for a density-dependent interaction effect. For example, a realistic scenario in this context could be the occurrence of three-body losses during the early loading phase of the lattice, where the scattering length is still significant \cite{Note1}. This inhomogeneous redistribution of entropy further highlights the relevance and necessity of our thermometry technique based on resolving individual atomic wires.

A direct follow-up study to the work presented here is the measurement of the equation of state of strongly interacting fermionic wires. Within the framework of the local density approximation, the precise knowledge of the axial potential will allow us to locally relate the 1D density and chemical potential at any interaction strength. The local pressure and compressibility can then be obtained from the integral and derivative of the density with respect to the chemical potential \cite{Tung2010,Ku2012,Fenech2016}. From these observables, further thermodynamic quantities can be determined, such as the reduced temperature $T/T_F$. Theoretically, the ground state and low-temperature regime ($T \ll T_{F}$) of the interacting 1D Fermi gas can be solved with the Bethe ansatz \cite{Guan2013a}. However, for the general finite-temperature case, the situation is significantly more challenging and theoretical studies are sparse \cite{Hoffman2015,Tajima2020}.

The individual probing of 1D Fermi gases promises further insight into elusive states of matter and critical behavior. This includes the observation of the highly sought FFLO phase \cite{Orso2007,Hu2007,Liao2010} and the study of TLLs featuring collectivized excitations and spin-charge separation \cite{Recati2003,Kollath2005,Yang2017,Yang2018,Vijayan2020}. Of particular interest is the interplay of strong interactions and the suppression of thermalizing collisions in 1D \cite{Kinoshita2006,Gring2012,Meinert2017}, which strongly impacts out-of-equilibrium and transport phenomena.

\begin{acknowledgments}
We thank \mbox{Tarik} \mbox{Yefsah} for stimulating and helpful discussions and critical reading of the manuscript as well as \mbox{Antoine} \mbox{Heidmann} for hosting the project at Laboratoire Kastler Brossel and his active support for the completion of this work. J.S. was supported by LabEX ENS-ICFP: ANR-10-LABX-0010/ANR-10-IDEX-0001-02 PSL*. We acknowledge support from DIM Sirteq (Grant EML 19002465 1DFG) and Fondation S. et C. del Duca (Grant 61846).

C.D. and M.T. conducted the measurements and data analysis. C.D. and J.S. performed the theoretical computations. C.D., M.T., C.E., T.R. and J.S. designed and constructed the relevant parts of the experimental apparatus for the project.  J.S. conceptualized, planned and supervised the project. C.D. and J.S. prepared the manuscript with comments from all authors.
\end{acknowledgments}


\bibliography{1D_FermiGas}

\end{document}


\pagebreak
\onecolumngrid
\begin{center}
\large{\textbf{Supplemental Material:\\\emph{In Situ} Thermometry of Fermionic Cold-Atom Quantum Wires}}
\end{center}
\twocolumngrid
\setcounter{equation}{0}
\setcounter{figure}{0}
\setcounter{table}{0}
\setcounter{page}{1}
\makeatletter
\renewcommand{\thepage}{S\arabic{page}} 
\renewcommand{\theequation}{S\arabic{equation}}
\renewcommand{\thefigure}{S\arabic{figure}}
\renewcommand{\bibnumfmt}[1]{[S#1]}
\renewcommand{\citenumfont}[1]{S#1}

\author{Cl\'{e}ment~De~Daniloff}
\affiliation{\afflkb}
\author{Marin~Tharrault}
\affiliation{\afflkb}
\author{C\'{e}dric~Enesa}
\affiliation{\afflkb}
\author{Christophe~Salomon}
\affiliation{\afflkb}
\author{Fr\'{e}d\'{e}ric~Chevy}
\affiliation{\afflkb}
\author{Thomas~Reimann}
\affiliation{\afflkb}
\author{Julian~Struck}
\email[Corresponding author:~]{julian.struck@lkb.ens.fr}
\affiliation{\afflkb}

\maketitle


\section{Large-spacing optical lattice}
\begin{figure}
\centering
\includegraphics[width=8.6cm]{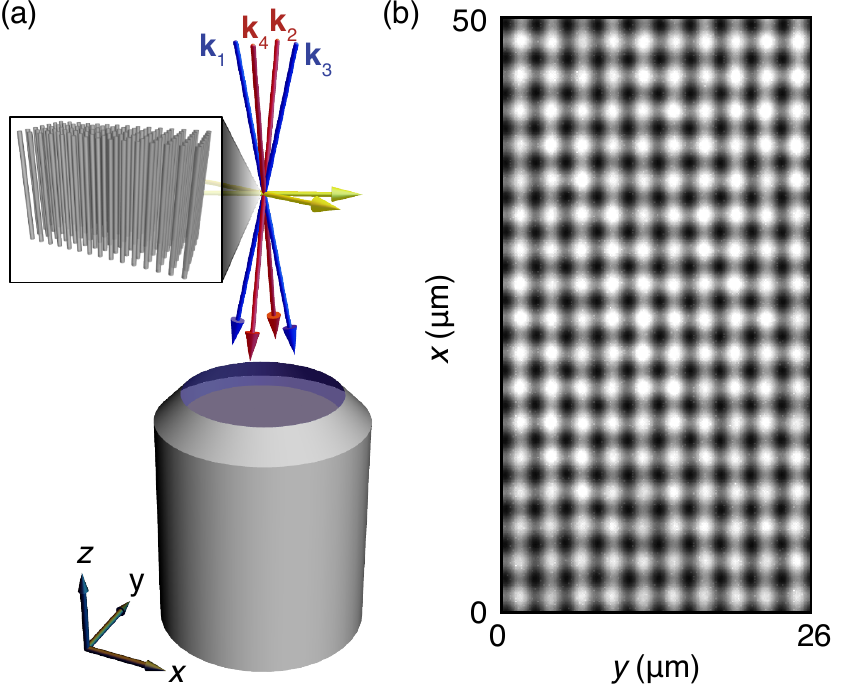}
\caption{Direct imaging of the optical lattice. (a) The laser beam pairs (blue and red) forming the lattice are collected by a microscope objective and (b) refocused onto a CCD camera. For the sake of completeness, the crossed optical dipole trap beams (yellow) are also included in (a).}
\label{fig:S1}
\end{figure}

The 2D optical lattice consists of two superimposed standing waves, which are each created by interfering a pair of laser beams under an angle of $2\theta\approx 24^{\circ}$. The standing waves along the $x$- and $y$-axis are formed by beam pairs with the wavevectors $\mathbf{k}_{1,3}=k\,(\pm\sin({\theta}),0,-\cos({\theta}))$ and $\mathbf{k}_{2,4}=k\,(0,\mp\sin({\theta}),-\cos({\theta}))$, respectively. Here, $k=2\pi/\lambda$ is the wavenumber and $\lambda=1064\,\mathrm{nm}$. The waists of the lattice beams are approximately $180\,\mathrm{\mu m}$, which results in a variation of $\omega_{\perp}$ of less than $2\%$ across the $18$ central tube traps that are used for data analysis. We collect all four lattice beams in a microscope objective after they have passed the atomic cloud and image them onto a CCD camera [see Fig. \ref{fig:S1}]. This gives us an absolute phase reference of the lattice. Through piezo-driven mirrors in the beam paths \cite{Magnan2018}, we control the phase and compensate long-term drifts at the beginning of each experimental sequence.

\begin{figure}
\centering
\includegraphics[width=8.6cm]{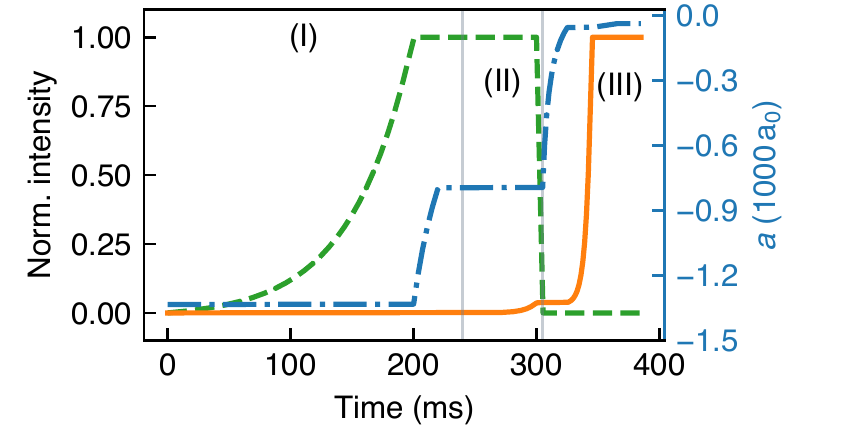}
\caption{Lattice loading ramps. (Green dashed line) Normalized intensity of the compression beam. (Orange solid line) Normalized intensity of the lattice laser beams. (Blue dashed-dotted line) $s$-wave scattering length. The roman numbers correspond to the schematic depiction in figure 2(a).}
\label{fig:S2}
\end{figure}

The intensity ramps of the compression beam and optical lattice for the preparation of a single row of tubes are shown in figure \ref{fig:S2}. Also depicted is the change of the $s$-wave scattering length during the loading procedure, starting from  $a=-1334\,a_{0}$ in the crossed optical dipole trap for the initial evaporative cooling of the atomic gas.

\section{Absorption imaging}
We resonantly image the energetically lowest hyperfine state $\ket{F=9/2,\,m_{F}=-9/2}$ at a magnetic bias field above $200\,\mathrm{G}$ on the D2 cycling transition. Prior to imaging, all optical potentials are switched off in order to avoid spatially dependent AC Stark energy shifts. A short imaging pulse of $t_{\mathrm{img}} = 12
\, \mathrm{\mu s}$ ensures that the gas expands only marginally in the transverse direction ($\omega_{\perp}t_{\mathrm{img}}/2\pi=0.2$) and that the frequency shift due to the Doppler effect is negligible. We have verified the latter by measuring the velocity of the atomic cloud after imaging pulses of different durations \cite{Hueck2017a}. 
A linear increase of the velocity with the pulse duration implies a constant force on the atoms during the imaging process and hence no significant frequency shift. The pulse duration of $12 \, \mathrm{\mu s}$ lies well within this regime.

\begin{figure}
\centering
\includegraphics[width=8.6cm]{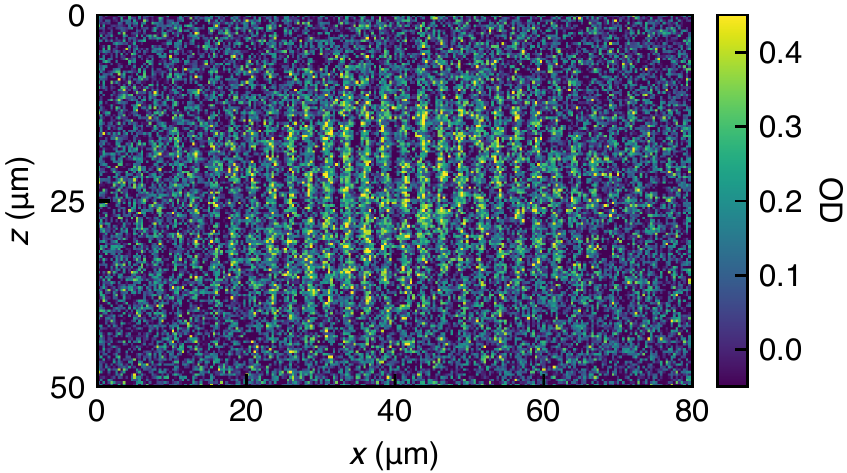}
\caption{Individual absorption image of atomic wires in the 1D regime. This image is part of the average of Fig. 1(b).}
\label{fig:SingleAbsorptionImage}
\end{figure}

To determine the column density $\tilde{n}$ of the atomic gas along the probe beam direction, we use the extended Beer-Lambert law
\begin{equation}
\sigma_{0} \tilde{n} = \mathrm{ln}(I_{\mathrm{ref}}/I_{\mathrm{abs}}) + (I_{\mathrm{ref}}-I_{\mathrm{abs}})/I_{\mathrm{sat}},
\label{eq:BeerLambert}
\end{equation}
which accounts for the saturation of the atomic transition. Here, $\sigma_{0}$ is the absorption cross section, $I_{\mathrm{sat}}$ the saturation intensity, $I_{\mathrm{ref}}$ and $I_{\mathrm{abs}}$ the imaging intensities before and after the atoms, respectively. Typically, we operate with a saturation parameter of $s=I_{\mathrm{ref}}/I_{\mathrm{sat}} \approx 3-4$. For its calibration, we prepare clouds of low atom number ($\mathrm{OD}<0.5$) under the same experimental conditions and vary the probe beam power. Subsequently, we find the saturation parameter that yields a constant column density [see eq. \eqref{eq:BeerLambert}] for varying probe beam powers. The estimated systematic uncertainty on the saturation parameter is $10\%$. Figure \ref{fig:SingleAbsorptionImage} shows an individual absorption image of atomic wires in the 1D regime.

\section{Optical pumping tomography}
\begin{figure}
\centering
\includegraphics[width=8.6cm]{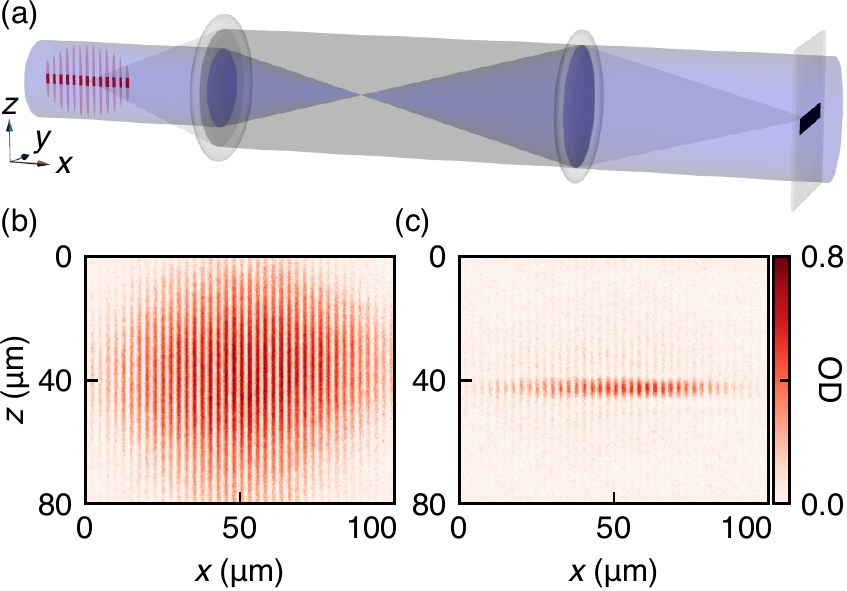}
\caption{Optical pumping tomography. (a) Imaging system projecting the shadow of an opaque slit onto the row of atomic wires. (b) and (c) correspond to absorption images of the atomic wires without and with optical pumping, respectively.}
\label{fig:S3}
\end{figure}

The large axial extent of the atomic wires poses a problem for the $z$-axis imaging system [see Fig. \ref{fig:S2}(a)], which features a depth of field of $\pm 5\, \mathrm{\mu m}$. We mitigate this issue by optically pumping atoms located outside of a central slice into the upper hyperfine ground state manifold ($m_{J}=1/2$), rendering them transparent to the probe beam. The pumping beam illuminates an opaque slit on an optical window, which is then projected onto the row of atomic wires [see Fig. \ref{fig:S3}]. The demagnified size of the slit at the position of the atoms is $6\,\mathrm{\mu m}$, while the resolution of the imaging system is $5\,\mathrm{\mu m}$. This is consistent with the measured  $4\,\mathrm{\mu m}$ full width at half maximum of the atomic slice after the tomography [see Fig. \ref{fig:S3}(c)]. In the experiment, the pumping pulse illuminates the atomic cloud for $1\,\mathrm{\mu s}$ before the absorption image is taken.

\section{Measurement of the Axial Frequency}
\begin{figure}
\centering
\includegraphics[width=8.6cm]{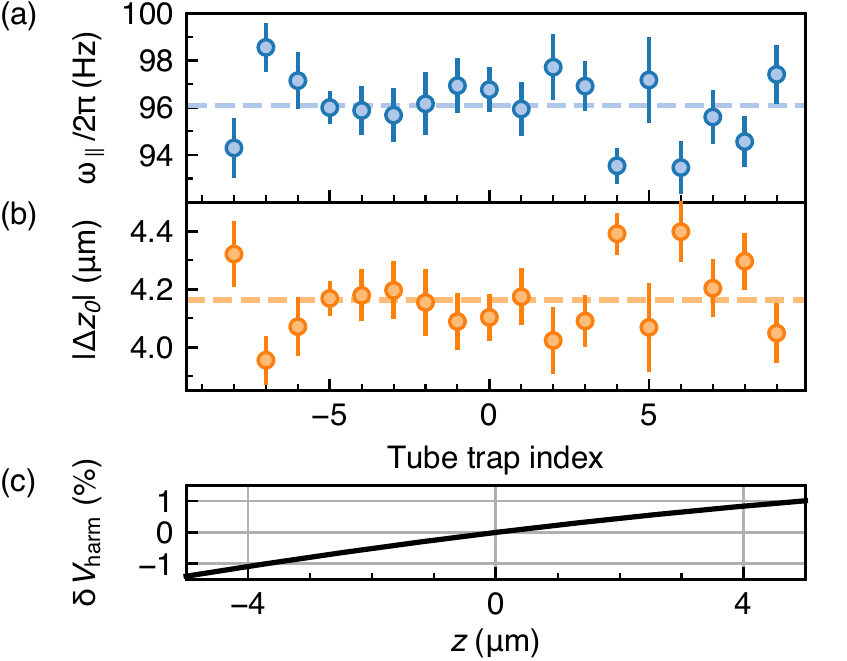}
\caption{Gradient measurement of the axial frequency. (a) Axial frequency of each tube trap. (b) Corresponding displacement $|\Delta z_{0}|$ of the maximum of the 1D density distribution. The dashed lines in (a) and (b) indicate the mean value over all tubes. (c) Relative difference $\delta V_{\mathrm{harm}}$ between the harmonic approximation and the fully reconstructed axial potential.}
\label{fig:S4}
\end{figure}
Due to the anharmonicity of the potential and the large axial extent of the atomic wires, a reliable measurement of the axial frequency via induced oscillations is not feasible. Instead, we apply calibrated magnetic field gradients $\pm B^{\prime}_{z}$ along the axial direction and measure the differential displacement $2\Delta z_0$ of the maximum of the 1D density distribution. In the small displacement limit ($\Delta z_0 \rightarrow 0$), the axial trapping frequency is given by $\omega_\parallel = \sqrt{\mu B^{\prime}_{z}/m \Delta z_0}$, where $\mu$ is the magnetic moment. 

We observe no statistically significant shift of the measured frequency across the 18 central tube traps [see Fig. \ref{fig:S4}]. Typically, the displacement from the center of the trap is $\Delta z_0 \approx \pm 4\,\mathrm{{\mu m}}$. A self-consistency check with the fully reconstructed potential demonstrates that the relative deviation from a harmonic potential in this region is at most $1\%$ [see Fig. \ref{fig:S4}(c)].

\section{Equation of State Fitting Procedure}
Using the local density approximation $\mu(z) = \mu_{0} - V_{\parallel}(z)$, with $\mu_{0} = \mu(z=0)$, we simultaneously fit a large number of 1D density profiles using the noninteracting equation of state $n(\mu(z),T)$ [Eq. (1)]. In order to account for the anharmonicity of the axial potential, we use a 7th order polynomial model 
\begin{equation}
V_\parallel(z) = \frac{1}{2} m \omega_\parallel^2 z^2 + \sum_{p=3}^7 a_p z^p,
\label{eq:AxialPotential}
\end{equation}
where the axial frequency $\omega_\parallel$ is directly measured as described in the previous section and the higher-order coefficients $a_p$ are left as fit parameters. We have verified that the reconstructed potential in the region with atoms is not changed for polynomial models of higher order than 7.

To prepare the 1D density profiles for the fitting procedure, we first sort all absorption images of a given data set according to their total atom number. Subsequently, this list is grouped in intervals of 10 images, which are then averaged. From the resulting images, the 1D density profiles for each tube are extracted. For example, the entire data set for the $s$-wave scattering length of $a=+40\,a_{0}$ consists of 270 absorption images. After averaging and selecting the 18 central tubes, we end up with a set of 486 individual 1D density profiles.

To measure $T$ and $\mu$ for a given profile and to determine the polynomial coefficients $a_{p}$ [Eq. \eqref{eq:AxialPotential}], we simultaneously fit an entire set of profiles. For each individual 1D profile, $\beta=1/k_{B}T$ and $\beta \mu$ constitute independent fit parameters, while the coefficients $a_{p}$ serve as shared parameters for the whole set. Statistical errors of the thermodynamic quantities are obtained from the covariance matrix of the fit.

\section{Transverse Mode Occupation}
\begin{figure}
\centering
\includegraphics[width=8.6cm]{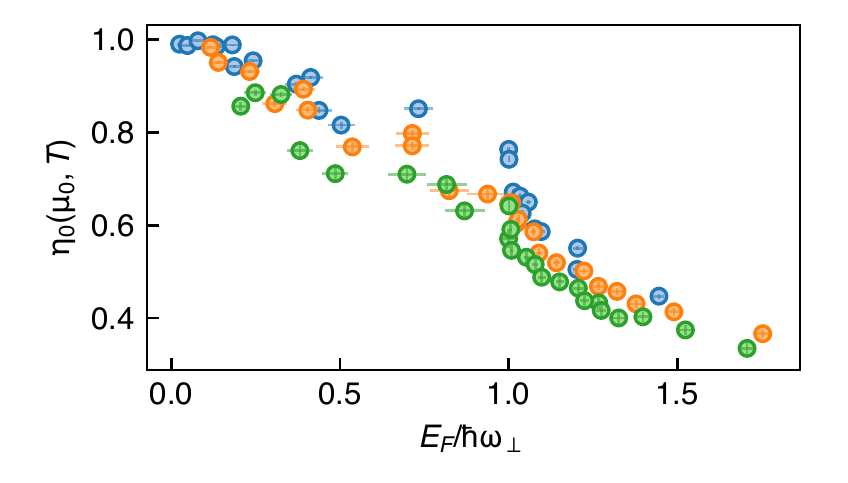}
\caption{Relative occupation $\eta_{0}$ of the lowest transverse mode at the center of the tube traps ($z=0$), where $\eta_{0}$ is the smallest. Different colors represent three characteristic tube traps. The data points correspond to the ones shown in Fig. 4.}
\label{fig:S5}
\end{figure}
The transverse mode population of the tube traps is a measure of the dimensional character of the atomic wires. The relative occupation of the lowest mode
\begin{equation}
	\eta_{0}(\mu,T) = \frac{\mathrm{Li}_{\frac{1}{2}}(-f_{0}(\mu,T))} {\sum_{s=0}^{\infty} (s+1)\, \mathrm{Li}_{\frac{1}{2}}(-f_{s}(\mu,T))},
	\label{eqn:GSoccupation}
\end{equation}
is given by the equation of state [Eq. (1)]. In the low atom number and temperature limit, $\eta_{0}$ saturates to one [see Fig. \ref{fig:S5}], which, by definition, corresponds to the 1D regime.

\section{Isentropic 1D-3D relation of the reduced temperature for the noninteracting Fermi gas}
\begin{figure}
\centering
\includegraphics[width=8.6cm]{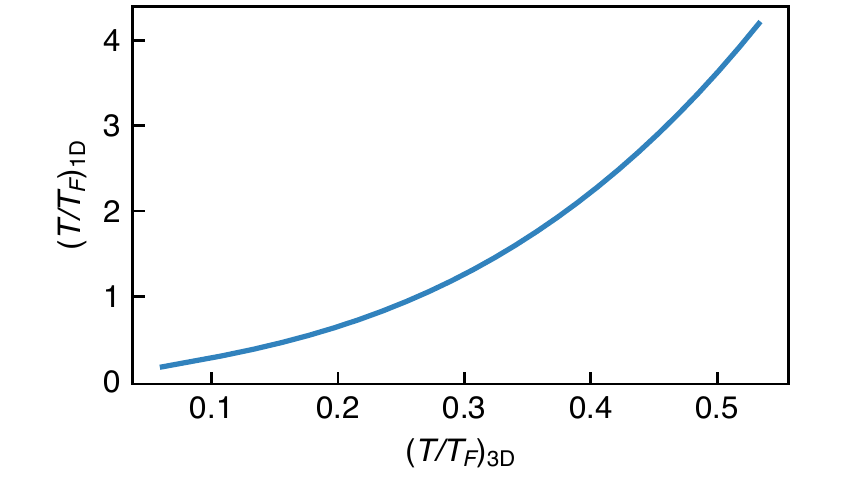}
\caption{Isentropic relation between the reduced temperatures of a homogeneous noninteracting 3D and 1D Fermi gas.}
\label{fig:S7}
\end{figure}

In this section we study the isentropic thermodynamic relation between the noninteracting homogeneous 3D and 1D Fermi gas. This elementary (toy) model already demonstrates that the reduced temperature is not conserved in the adiabatic 1D-3D crossover. For the 3D and 1D Fermi gas, the corresponding reduced temperatures are given by
\begin{equation}
	\left(\frac{T}{T_{F}}\right)_{\mathrm{3D}} = 4\pi \left(-6\pi^2\,\mathrm{Li}_{\frac{3}{2}}(-f_{\mathrm{3D}	})\right)^{-2/3}\,,
	\label{eqn:ReducedT3D}
\end{equation}
\begin{equation}
	\left(\frac{T}{T_{F}}\right)_{\mathrm{1D}} = \frac{4}{\pi} \left(\,\mathrm{Li}_{\frac{1}{2}}(-f_{\mathrm{1D}})\right)^{-2}\,,
	\label{eqn:ReducedT1D}
\end{equation}
and the entropy per particle reads
\begin{equation}
	\left(\frac{S}{N k_{B}}\right)_{\mathrm{3D}} = \frac{5 \, \mathrm{Li}_{\frac{5}{2}}(-f_{\mathrm{3D}	})}{2 \, \mathrm{Li}_{\frac{3}{2}}(-f_{\mathrm{3D}})} - \ln(f_{\mathrm{3D}})\,,
	\label{eqn:Entropy3D}
\end{equation}
\begin{equation}
	\left(\frac{S}{N k_{B}}\right)_{\mathrm{1D}} = \frac{3 \, \mathrm{Li}_{\frac{3}{2}}(-f_{\mathrm{1D}})}{2 \, \mathrm{Li}_{\frac{1}{2}}(-f_{\mathrm{1D}})} - \ln(f_{\mathrm{1D}})\,,
	\label{eqn:Entropy1D}
\end{equation}
where $f_{\mathrm{3D}}$ and $f_{\mathrm{1D}}$ are the fugacities of the 3D and 1D gas, respectively. Since the entropy is conserved, i.e. $\left({S}/{N k_{B}}\right)_{\mathrm{1D}} = \left({S}/{N k_{B}}\right)_{\mathrm{3D}}$, $f_{\mathrm{3D}}$ and $f_{\mathrm{1D}}$ are directly connected through Eqs. \eqref{eqn:Entropy3D} and \eqref{eqn:Entropy1D}. With the isentropic relation between the fugacities we then obtain the isentropic relation of the reduced temperatures in 3D and 1D (see Eqs. \eqref{eqn:ReducedT3D}, \eqref{eqn:ReducedT1D} and Fig. \ref{fig:S7}).

\bibliography{SM_1D_FermiGas}